# Understanding grain boundary electrical resistivity in Cu: the effect of boundary structure

Hanna Bishara [(a)]*, Subin Lee [(a)(b)], Tobias Brink [(a)], Matteo Ghidelli [(a)(c)], Gerhard Dehm [(a)]*

[(a)] Max-Planck-Institut für Eisenforschung GmbH, 40237 Düsseldorf, Germany

[(b)] Institute for Applied Materials (IAM), Karlsruhe Institute of Technology, 76344 Eggenstein-Leopoldshafen, Germany

[(c)] Laboratoire des Sciences des Procédés et des Matériaux (LSPM), CNRS, Université Sorbonne Paris Nord, 93430 Villetaneuse, France

* Corresponding author: h.bishara@mpie.de ; g.dehm@mpie.de

## Abstract

Grain boundaries (GBs) in metals usually increase electrical resistivity due to their distinct atomic arrangement compared to the grain interior. While the GB structure has a crucial influence on the electrical properties, its relationship with resistivity is poorly understood. Here, we perform a systematic study on the resistivity–structure relationship in Cu tilt GBs, employing high-resolution in-situ electrical measurements coupled with atomic structure analysis of the GBs. Excess volume and energies of selected GBs are calculated using molecular dynamics simulations. We find a consistent relation between the coincidence site lattice (CSL) type of the GB and its resistivity. The most resistive GBs are high range of low-angle GBs (14°–18°) with twice the resistivity of high angle tilt GBs, due to the high dislocation density and corresponding strain fields. Regarding the atomistic structure, GB resistivity approximately correlates with the GB excess volume. Moreover, we show that GB curvature increases resistivity by ~80%, while phase variations and defects within the same CSL type do not considerably change it.

**Introduction**

The electrical resistivity of grain boundaries (GBs) in conductive materials hampers the development of the newest generation of nano-electronic and energy-harvesting devices. For instance, GB resistivity is a major concern for electron transport in sub-20-nm interconnects in integrated circuits, [1–3] while GBs in thermoelectric and photovoltaic materials are suspected to decrease device efficiency. [4,5] To overcome these challenges, GB engineering has been utilized to optimize the material's functional performance, for example through a controlled sample preparation or processing. [6–8] In doing so, the scientific community relies on experimental evidence that GB resistivity is decreased for low-angle GBs (LAGBs) and twin boundaries. [3,6,7,9,10] Nevertheless, experimental studies on the impact of GB characteristics, such as, type, misorientation, or inclination, phase, and curvature on resistivity are still missing due to limitations of the spatial resolution and sensitivity of resistivity measurements.

The relationship between the resistivity of a GB and its structural characteristics arises from the altered atomic structure of the GB compared to the grain interior. This creates a fluctuation in the periodic atomic potential from the adjacent crystals across the boundary, leading to electron scattering at the boundary by a potential wall. The magnitude of the potential wall is associated with the GB structure and its chemical bonding. [11,12] Moreover, the distinct atomic arrangements at the GB also locally change the density of states and electron density compared to the grain interior [13] as confirmed by density functional theory (DFT) simulations. [10,12,14] However, its experimental observation is challenging because of the difficulties in isolating a specific GB and characterizing solely its resistivity. [11,15,16] Hence, cumulative scattering events on the different GBs blur out all details of the influence of GB type and character on resistivity. To overcome this challenge, there is a need to probe the electrical resistivity of an individual GB segment. Nakamichi [17] inspected individual GBs in bulk bicrystals and experimentally revealed a misorientation dependence of the resistivity. However, this study, which was conducted at cryogenic temperature, did not consider the GB characteristics, e.g. its inclination, phases, curvature or defects. Later, small-scale approaches were introduced to locally probe single, sub-micron GB segments using micromanipulators. [18–20] Recently, we further improved this method to gain ultra-high sensitivity, enabling resistivity measurements of a Σ3 Cu GB. [21] Here, we adopt this technique and extend the research to a systematic study of the effect of different individual GB structures on resistivity in a polycrystalline Cu thin film with [111] tilt GBs.

Within the context of tilt GBs, the geometric relation between neighbouring grains is described through the tilt axis, misorientation angle $\theta$, and the normal(s) of the GB plane(s) in each case. For symmetric GBs, a common plane exists, while for an asymmetric boundary the GB plane normals are different in the adjacent grains. For discrete tilt angles $\theta_{CSL}$, corresponding to coincidence site lattices (CSLs), GBs exhibit unique periodic atomic structures (motifs). [22] The same motifs prevail for GBs even at some deviations of $\delta\theta$ from the exact $\theta_{CSL}$ within the Brandon criterion (maximum angle of deviation from an exact CSL that could be sustained by a dislocation array).[23] The atomic configurations of GBs can be predicted by atomistic simulations and resolved through the aberration-corrected (scanning) transmission electron microscopy ((S)TEM) imaging.[24–26] The equilibrium structure of the GB in a pure material depends not only on the misorientation between the neighbouring grains and the GB plane, but also on temperature and pressure conditions. Analogous to bulk phases, the resulting structures are referred to as GB phases.[24,27,28] For metals, it has only recently been observed that a specific GB can have different phases[24,29] as predicted earlier by interface thermodynamics.[27,28,30] In terms of interface thermodynamics GB phases are described by their excess interfacial energy $E_{gb}$, excess volume $\Delta V$, excess entropy, and interface stress.[30] These thermodynamic state variables determine the GB phase stability, while kinetics control the transformation velocities. In recent studies, three GB phases were observed for Cu Σ19b GBs via atomic-scale STEM and by molecular dynamics (MD) computer simulations.[29,31] In some cases, different GB phases were found in the same GB segment.

In this context, this study focuses on Cu, on the one hand as a model system that has been investigated with respect to different GB structures, and on the other hand due to the high application relevance concerning its electronic properties for integrated circuits. We consider different Cu [111] tilt GBs, namely Σ3, Σ7, Σ19b, Σ21a, and Σ37c, as well as LAGBs with misorientation angles ranging from 7° to 18°. The study sheds light on similarities and differences in GB resistivity for (*i*) symmetric and asymmetric variants, (*ii*) variations in GB inclination, (*iii*) deviations from the ideal CSL, (*iv*) introduction of twist components, (*v*) macroscopic curvature, and (*vi*) possible different GB phases (see Fig. 1).

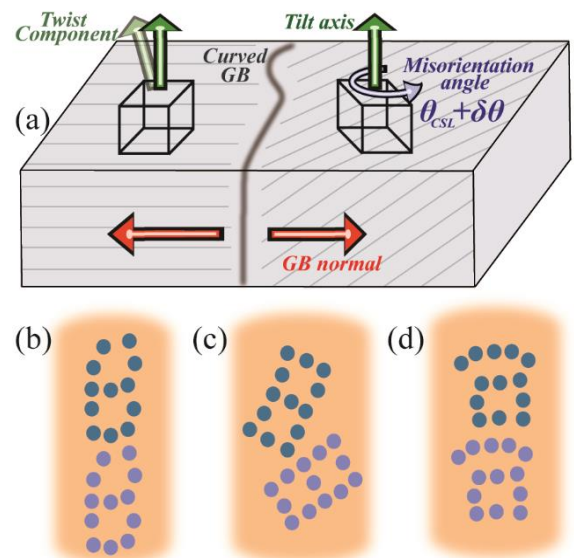

Figure 1: (a) Parameters to describe the relative crystallographic orientation between adjacent grains. The <111> tilt axis is indicated by green arrows. An inclination of this tilt axis leads to a twist component within the GB. GB plane normals are represented by red arrows. The resulting misorientation between the grains ($\theta_{CSL} + \delta\theta$) determines the CSL value, with the deviation from the exact CSL angle denoted as $\delta\theta$. Finally, GBs might be straight or curved macroscopically. For Σ19b GBs ($\theta = 46.8°$), three possible GB phases occur with different atomic arrangements [29,31]: (b) zipper for the symmetric $(23\bar{5})$ GB plane, as well as (c) domino and (d) pearl for the symmetric $(1\bar{8}7)$ GB plane. The atomic arrangements are shown here in the projection from the <111> direction.

## Results and discussion

### Selection of different grain boundaries

Well-defined tilt GBs in Cu are achieved through the deposition of a thin film by magnetron sputtering on a c-plane α-Al$_2$O$_3$ (sapphire) surface, as this is known to create [111] tilt GBs aligned vertical to the surface. [32,33] The electron backscatter diffraction (EBSD)-resolved inverse pole figure maps, shown in Fig. 2a,b, confirm an abnormal grain growth and a uniform (111) planes parallel to the surface. Additionally, the subtly different blue colours in Fig. 2a indicate a small twist component between some of the grains up to 3°. The (111) pole figure in Fig. 2c implies two ranges of misorientation angles: below ~21°, and between 42° and 60°. The majority of the GBs is Σ3 boundaries and there are CSL boundaries also, for example Σ7, Σ19b, Σ21a, and Σ37c. Among them, some GB segments are selected for the resistivity measurements. The GB characteristics of each segment are evaluated by scanning electron microscope (SEM) and EBSD analysis of its adjacent grains (Fig. 2d) as described in the Methods section.

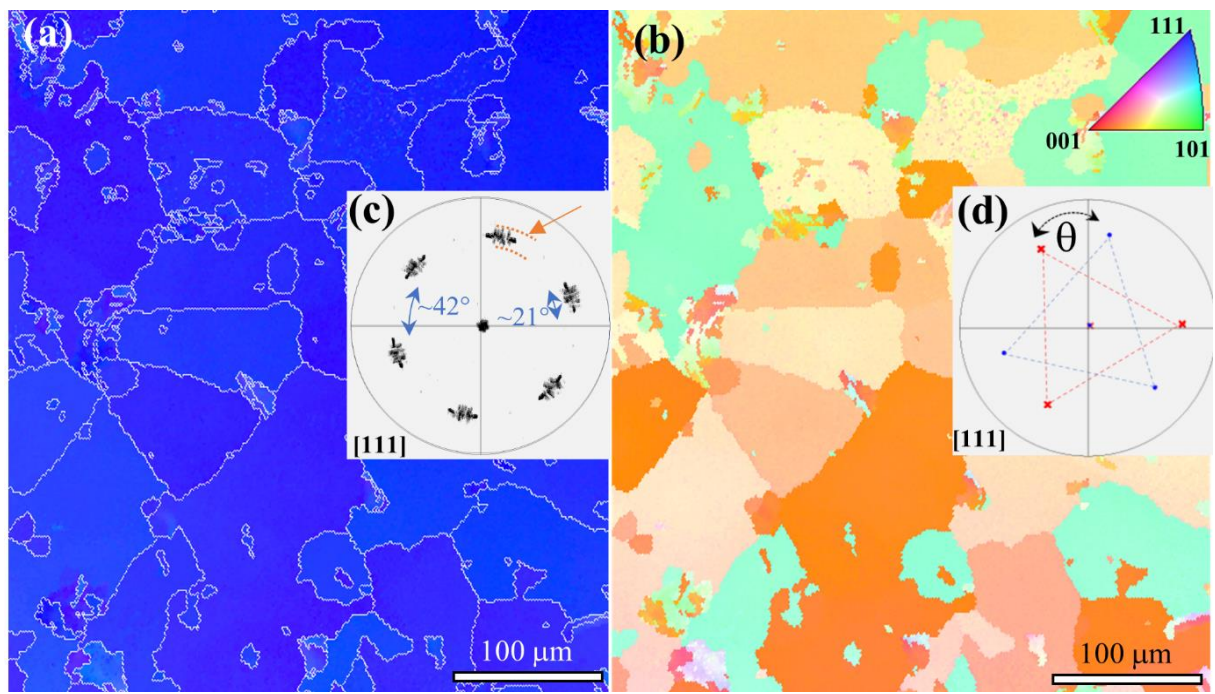

Figure 2: [111] inverse pole figure EBSD-resolved map for (a) out-of-plane and (b) in-plane orientations of the annealed Cu thin films. The maps indicate abnormal grain growth and preferred crystallographic orientation. (c) Pole figure showing a range of [111] tilt grain boundary misorientation. The radial width of the reflection is the maximum spread of twist (orange mark) (d) Pole figure of a selected bicrystal utilized to identify the tilt and twist component of the GB.

The selected GB segments are isolated from their surrounding material by focused ion beam (FIB) milling of two trenches across the film to create a conduction line that includes the GB. Inside the SEM, four needles with 50 nm tip radius probe the FIB-milled structures by forming electrical contacts across the GB as shown in Fig. 3. The position of the needles is accurately controlled by piezo-driven micromanipulators. Current is applied through the outer needles (marked by #1 and #4 in Fig. 3),, while voltage is measured between the inner needles (marked by #2 and #3 in Fig. 3), which are separated by distance $L$. The resistance $R$ for a conduction line is given by [21]

$$(1) \qquad R = \frac{L}{A_{\text{line}}}\rho + \frac{\gamma}{A_{\text{GB}}},$$

where the first term is the contribution of the grain interior with bulk resistivity $\rho$. The second term is the GB contribution to resistance through GB resistivity $\gamma$. This term affects the resistance only when the voltage drop is measured across the GB. $A_{\text{line}}$ and $A_{\text{GB}}$ represent the cross sections of the conduction line and GB. The measurement technique and its reliability, as well as the use of equation 1 to extract GB resistivity are described in detail in Ref. [21].

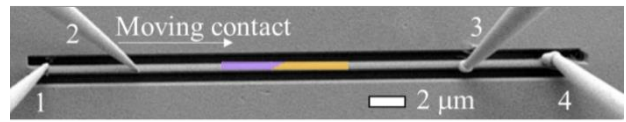

Figure 3: Electrical measurements across a GB. Current is supplied by needles 1 and 4, voltage is measured through needles 2 and 3. All needles are fixed at constant position except needle 2, which scans across the boundary. The EBSD-resolved grain map is laid on top of the GB region. (The tips of needles 3 and 4 are reshaped after electrostatic de-charging while positioning them inside the SEM.)

**Dependence of GB resistivity on CSL**

GBs with different misorientations were isolated and electrically measured. The GB resistivity dependence on the tilt misorientation (Fig. 4) exhibits two regimes; LAGBs where the resistivity increases with the tilt angle, and high angle regime which shows smaller resistivity with larger tilt angle. It suggests that the scattering potential barrier of a GB correlates with its CSL type, while the barrier is only slightly affected by the GB structural characteristics within the same Σ-type. The range of GB resistivity (vertical length of the coloured rectangles) covers the spread from all measured segments within the same CSL type, as well as the error bars obtained from 2–3 repeating electrical measurements on the same GB segment. Multiple segments of the same Σ-type boundary are distinguished by structural variations, which do not noticeably affect resistivity.

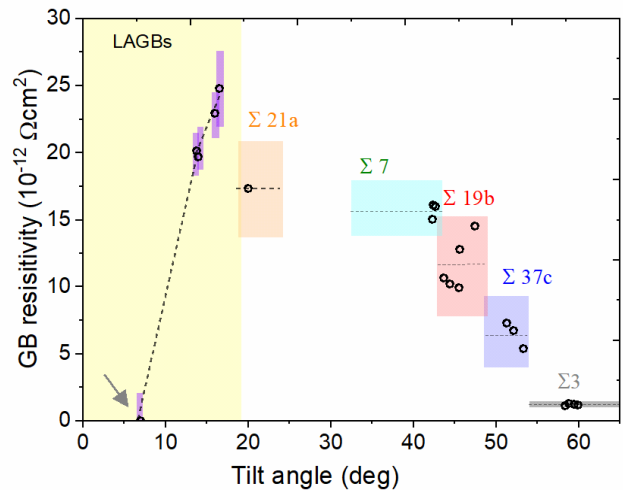

Figure 4: GB resistivities of different CSL GBs and LAGBs. The coloured regions represent the spread of the measured values. In the vertical direction they include the resistance measurement deviation obtained from all measurements within the same CSL type (including different GB structural characteristics). In the horizontal direction, they represent the angular tolerance of CSL type set by the Brandon criterion. The dashed line indicates the average value of resistivity for each CSL type. The resistivity of a 7° LAGB (marked with an arrow) is below the sensitivity of the measurement method.

Theories on GB resistivities, e.g. Mayadas–Shatzkes theory and its extensions, [11,34] relate the strength of the scattering potential at the boundary to the loss of atomic periodicity of the crystal and a change in Fermi velocity of electrons propagating in different directions. However, these models do not supply further information on the effect of characteristics of the material and the boundary (e.g. GB misorientation or motif/phase) on resistivity. On the other hand, DFT calculations predicted that GB resistivity values are a function of the interface excess energy $E_{gb}$. [12] Indeed, from a structural point of view, quantification of the altered atomic structure at the boundary compared to the grain interior is made through interface excess properties, e.g. interface excess energy ($E_{gb}$) and excess volume ($\Delta V$). Therefore, we assume that $E_{gb}$ and $\Delta V$ represent the deviation of a GB from the background crystalline potential. An increase in excess properties leads to an increase in the fluctuating atomic potential of a GB relative to the bulk, and consequently a higher scattering potential. [11]

To explore such a correlation between GB resistivity and its excess properties, we searched for GB structures with MD annealing simulations using an embedded atom method (EAM) potential [35] that has been successful in reproducing experimental GB structures. [26,29] For each of Σ21a, Σ7, Σ19b, and Σ3, we used both possible symmetric GB planes. The resulting structures are shown in Supplemental Fig. S1 and the excess properties are plotted in Fig. 5a. In case of Σ21a {123}, two structures are observed, one of which resembled a dense dislocation structure with relatively disordered cores. The atomic structure of Σ21 {145} GB clearly consists of a dislocation array and resembles an LAGB. For Σ19b {178}, we also included the metastable "domino" structure, which was also found to occur in experiment.[29] The structure of Σ19b {235} matches to earlier experimental work.[31] The Σ3 {110} data is included, but it should be noted that this GB is prone to faceting, as also observed in earlier simulations.[36,37] While these structures do not represent the complete multitude of asymmetric and defective GB segments which are present in the experimental samples, we can use this data to evaluate general correlations between excess properties and resistivity.

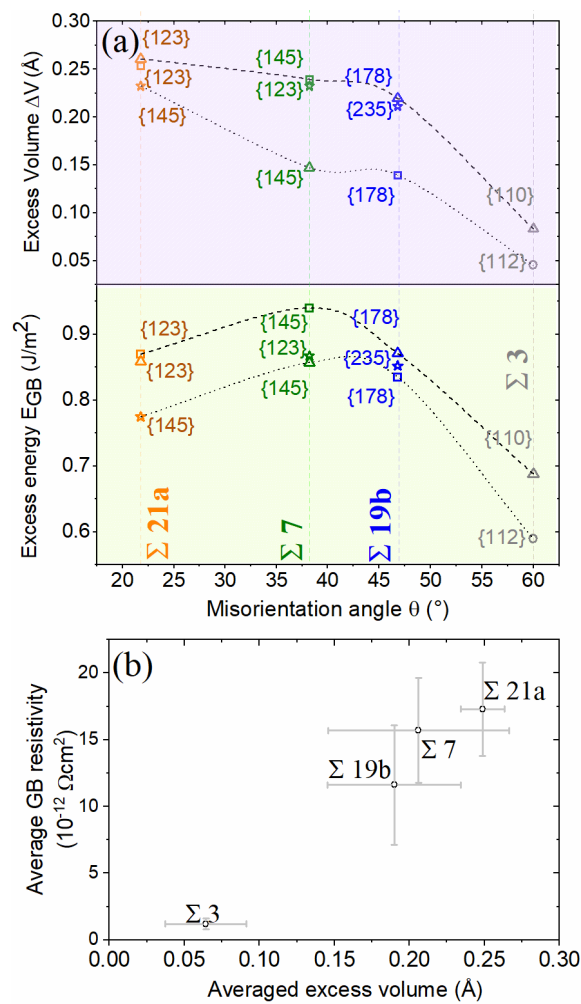

Figure 5: (a) Excess volume and excess energies for several GB phases as calculated using MD simulations. Dashed and dotted lines connect the higher and lower excess values for each CSL type, as a guide for eyes. Each point corresponds to the indicated GB planes, which in some of the cases have multiple possible phases. (b) Correlation between average GB resistivity and averaged excess volume for Cu CSL GBs. The error bars in horizontal direction cover the range of excess volumes of the different GB phases and correspond to Fig. 4 in vertical direction.

A comparison between GB resistivities and GB excess properties (Figs. 4 and 5), which were calculated based on method in Ref [38], leads to two main observations. First, resistivity is approximately proportional to the excess volume (Fig. 5b), even when considering the substantial variation of $\Delta V$ between different GB phases for the same CSL type. The correlation with $E_{gb}$ is weaker, with Σ21a having a lower excess energy, but higher excess volume than the other inspected GBs. This can be explained by its partial LAGB character as deduced from our simulations. Second, there is no obvious correlation between excess properties and resistivity within a specific Σ-type GB. For example, Fig. 5 shows that the differences in $E_{gb}$ and $\Delta V$ of the Σ19b GB phases can be larger than the differences of the average values of Σ19b and Σ7 GBs. However, such a difference is not reflected in the resistivity, as seen in Fig. 4. This could mean that, despite $E_{gb}$ and $\Delta V$ being clearly connected to GB resistivity across a set of misorientation angles, the effects of structure for a given Σ value are weak or dominated by defects in the experimental GBs, which will be discussed in more detail in a subsequent section. One should keep in mind that two phases can appear simultaneously within the same GB segment and consequently most of the electrical current would go through the less resistive phase with corresponding lower $\Delta V$. Secondary effects related to the electronic structure at the boundary might also affect resistivity. [13,39] The electronic relations require a separate study, as the current work focuses on the structural effects. Nevertheless, it is worth noting that the reproducibility of the scattering potential for the selected GBs provides confidence for designing and engineering GBs with predictable resistivities.

**Resistivity of low angle GBs**

The distinct structure of LAGBs from the CSL GBs necessitates a different approach to understand the angular misorientation dependence of GB resistivity. While the high-angle CSL GBs consist of atomistic motifs as repeating units, LAGBs with misorientation $\theta$ consist of an array of dislocations aligned along the boundary, separated by a distance $d = (b/2) \cdot \sin(\theta/2)$, where $b$ is the Burgers vector. The resistivity of LAGBs with $\theta = 7°$ (±1°) is below the detection limit of the experimental setup. However, an increasing misorientation angle is accompanied by a pronounced monotonic resistivity increment to values which exceed high angle GBs resistivities, as reported in Fig. 4. Resistivities of LABGs with more than 10° misorientation are the highest among the measured values, in agreement with earlier predictions that were based on the densities of dislocation arrays. [9]

To understand the resistivity–misorientation angle relation in LAGBs, the dislocation periodicity is examined. Dislocations within the 7° LAGB are expected to have a spacing of ~2.1 nm under the assumption that the Burgers vector is of 1/2 <110> type as in face-centred cubic dislocations. [40] The line density of dislocations created by the 7° misorientation does not noticeably affect resistivity as shown in Fig 4. However, an increase in the misorientation to 14° and 18° LAGBs yields a shorter inter-dislocation distance of 1.05 nm and 0.82 nm, according to the LAGB Read–Shockley model. [41] STEM high-angle annular dark field (HAADF) images of the 14° boundary show that edge dislocations are aligned along the grain boundary with a separation distance of ~1 nm (Fig. 6), which matches well with the aforementioned calculation. From the strain map calculated by geometrical phase analysis (GPA) and fast Fourier transformation analysis, it is found that the 1/2 <110> dislocations dissociate into two 1/6 <211> partials, as observed in Ref. [40]. The distance between a set of partial dislocations is comparable to the distance calculated with the Read–Shockley model for full dislocations. This creates a severely strained dislocation array (Fig. 6c), in agreement with the deviation from linear elasticity of the Read–Shockley equation for the higher-angle LAGBs. [41] Therefore, the high resistivities of the 14° and 18° GBs are attributed to the highly dense dislocation arrays as well as to the dislocation-induced stress (and strain) field. [41] Although the GB with 18° misorientation was not characterized by TEM, its electrical behaviour suggests that it follows the LAGB behaviour and consists of an array of dislocations. [41] A further increase in misorientation to ~21° results in a resistivity drop, i.e. the resistivity does not follow the increasing LAGB energy but a relaxed energy of a CSL structure (Σ21a). [22,42]

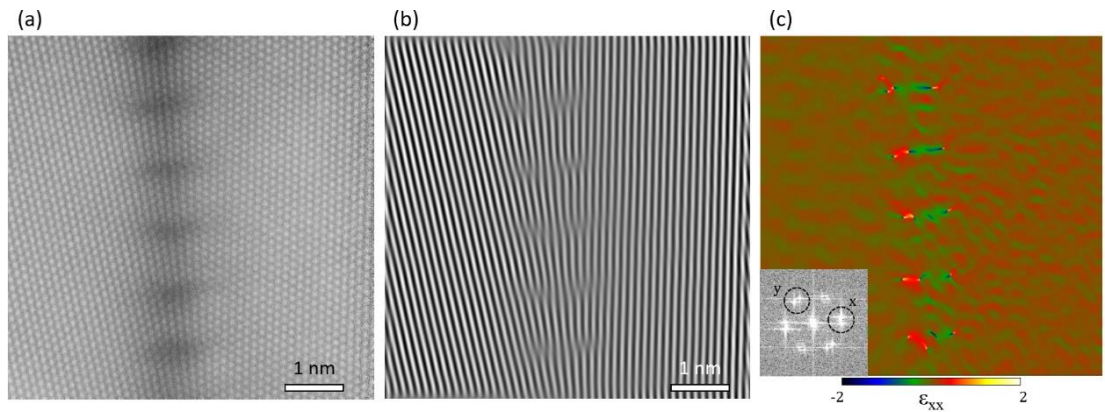

Figure 6: Dislocation array at the 14° LAGB. (a) STEM-HAADF image of the grain boundary. (b) Fast Fourier transform (FFT)-filtered image of (a). Only (111) lattice fringes are shown to visualize the extra half planes. (c) Strain map in the dislocation array analyzed by GPA. FFT in the inset shows the defined direction of the strain. There is a strong strain field at the grain boundary, especially near the core of the partial dislocations.

**GB resistivity within same CSL**

The effect of structural variation (i.e. inclination, twist, GB plane normal) within the same CSL type on the resistivity is investigated. This is also required to ensure that the comparison between the different Σ-type boundaries is reliable. In this context, the electrical resistivities of Σ19b GB segments having structural variations are acquired through direct resistivity measurements (Table 1). From earlier atomic resolved STEM studies, different GB phases are known to exist for Σ19b [111] $(23\bar{5})$ and $(1\bar{8}7)$ GB plane normals as illustrated in Fig. 1b-d. [29,31] It is worth noting that the GB planes extrapolated from the EBSD data refer to an approximation of a straight boundary line over 500–800 nm (width of the cut conduction line), i.e. they represent an equivalent straight GB with the corresponding average plane normals. The measured resistivity is similar, within the measurement errors, for all the inspected segments – despite the variation in the GB characteristics.

Table 1: Measured resistivity for Σ19b GBs. The segments are distinguished by symmetry or asymmetry, twist component, GB planes, and deviation from ideal CSL angle. GB phases are based on Refs. [29,31]. Resistivity is not noticeably affected by these variations within the same CSL type.

| GB type | Twist component | δθ (°) ±0.5 | GB planes | GB phase | GB resistivity ($10^{-12}$ $\Omega cm^2$) |
|---|---|---|---|---|---|
| **Sym.** | Yes | 1.2 | $[23\bar{5}]$ | Zipper | 12.8 ± 1.1 |
| **Sym.** | No | 0.8 | $[1\bar{8}7]$ | Pearl-Domino | 14.5 ± 3.5 |
| **Asym.** | No | 2.5 | $[2\bar{3}\bar{1}]$; $[9\bar{5}\bar{4}]$ | Zipper | 9.9 ± 1.7 |
| **Asym.** | Yes | 2.4 | $[10\bar{1}]$; $[4\bar{3}\bar{1}]$ | N.A. | 10.2 ± 2.5 |
| **Asym.** | No | 3 | $[3\bar{1}\bar{4}]$; $[2\bar{1}\bar{1}]$ | N.A. | 10.4 ± 1.2 |

The similar resistivities for the different Σ19b GB segments imply that the scattering potential of a boundary does not noticeably change within the same CSL GB type. While it is possible that the different GB phases simply have very similar resistivities, it is more likely that the average GB resistivity is a result of the imperfect GB structures in real materials, which over distances of several hundred nanometres contain multiple deviations from the ideal GB motifs to compensate for local twist components and inclination changes. For instance, the symmetric GB segments (Table 1) with zipper and pearl/domino structures (Fig. 1 b-d) exhibit similar resistivity, while both similarly deviate from an ideal CSL condition. This can explain the unchanged resistivity values for the asymmetric GB segments, since they tend to decompose into symmetric facets and steps in case of the zipper structure, while for the pearl structure additional sub-units are incorporated to compensate for the asymmetric inclinations according to previous TEM studies. [31] Such decomposition definitely creates variations in the atomic structure having a higher scattering potential than the ideal GB unit. Additionally, variations in the GB character, such as a 3° twist component and inclination, still result in similar resistivity values.

Apart from a periodic atomic ordering, a GB consists of defects such as disconnections and dislocations. The defects may appear for several reasons: compensation of GB inclination, asymmetric boundary dissociation, and phase transitions where two phases are separated by a line defect [31,43] (the macroscopic curvature of the GB does not belong to this group in the current context). Such imperfections do not noticeably change resistivity within a specific CSL tilt GB as indicated by the similar values for the asymmetric and symmetric segments, where the former are expected to contain a higher defect density than the latter, since they can decompose into symmetric segments. [31] Similar results (not presented here) are obtained for the other investigated Σ-types GBs.

**Absolute values of GB resistivity**

The measured resistivities of the GBs ($1-30\cdot10^{-12}$ $\Omega cm^2$) are higher by one order of magnitude than the values reported for Cu by macroscopic measurements and predicted by simulations ($0.1-4\cdot10^{-12}$ $\Omega cm^2$). [6,12,14,15,44,45] This might arise from the way GB resistivity values are simulated with DFT, where a relatively low amount of atoms in a defect-free periodic structure is considered, while real GB structures are never defect free. [12,14,29,46] Consequently, the calculated values only give lower bounds for the GB resistivity. In addition, our findings also overestimate the resistivity compared to macro-scale experiments. This difference could arise due major limitations of the macroscopic resistivity model, such as not considering GBs aligned along the electric field direction, deviation of scattered electrons from planar wave functions, and varying strengths of potential walls at different GB types. These challenges are comprehensively described in Ref. [34].

Despite the high absolute resistivity values measured in this experiment compared to literature, it is still possible to compare relative resistivities of GB types based on predictions. The lack of DFT simulations on resistivities of the investigated GBs, except for Σ3, [12,14,46] prevents us from a direct comparison between our experimental findings and theoretical predictions. Our measured resistivity values of incoherent Σ3 GBs (Fig. 4) are higher by almost an order of magnitude than the simulated values for a coherent boundary, $1.06\cdot10^{-12}$ vs. $0.2\cdot10^{-12}$ $\Omega cm^2$. This difference may be attributed to the different coherency of the GB, where the former is measured for incoherent boundary and the latter calculated for a coherent GB. Calculations predict that high-angle coherent and symmetric CSL GBs have a 10–20 times higher resistance than the coherent Σ3 twin boundary depending on the GB type. [12,14] This result matches our experiments, as the resistivity of high-angle CSL GBs is more than an order of magnitude higher than the resistivity of Σ3 GBs (Fig. 4). A similar comparison also applies to LAGB resistivities, where the resistivity values are negligible in the low-angle range, but more pronounced with increasing misorientation angle. [9]

**Effect of GB curvature**

GBs usually tend to curve instead of following a straight line, especially for the nano-grains (nanocrystals) found in confined integrated circuits. So, understanding the effect of the boundaries' curvature on resistivity is of high significance. To inspect this effect, macroscopically straight and macroscopically curved GB segments are isolated as shown in Fig. 7a and Fig. 7b, respectively. The straight GB segments are described by a pair of GB planes (or a single GB plane for the symmetric GBs) for all the GBs discussed in Fig. 4 and Table 1. However, curved GB segments cannot be described in this way, due to the changing inclination along the curve. A significant increase in GB resistivity is observed for GBs with macroscopic in-plane curvature relative to macroscopically straight GB segments. Specifically, curved GBs with misorientations of 14° and 18° exhibit resistivities of (36.1 ± 6) and (40.3 ± 6) ·$10^{-12}$ $\Omega cm^2$, compared to (19.7 ± 1.2) and (22.9 ± 1.7) ·$10^{-12}$ $\Omega cm^2$ in straight segments, respectively.

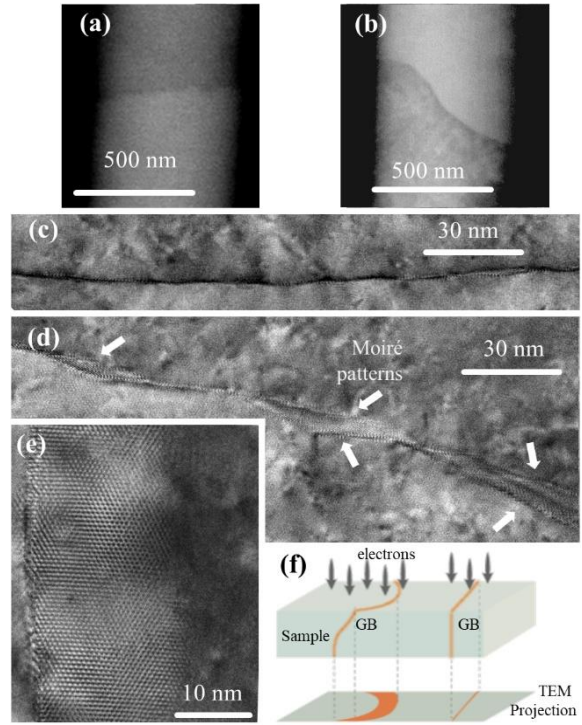

Figure 7: Top view SEM images using BSE detector for (a) straight and (b) curved 14° LAGBs. Top-view high-resolution TEM aberration corrected images of (c) the straight and (d) the curved GBs showing Moiré fringes for the curved GB segment. (e) Higher magnification of the Moiré pattern observed for the curved GB segments, revealing an inclination along the TEM imaging direction. (f) Illustration of the TEM projection for straight and laterally curved boundaries.

To analyse the structural differences between the straight and curved segments, a 14° misorientation LAGB has been observed by top-view TEM. At an atomic scale, both straight and curved GB segments have a similar atomic structure consisting of an array of edge dislocations (Fig. 6). Obviously, the GB defect density (e.g. disconnections) in the vicinity of a curved GB area is higher compared to the straight segments. However, it has already been shown that the defect density does not noticeably alter resistivity (Table 1). Yet, as the GB plane normals are do not to affect resistivity, then the curve must play a role in setting the GB resistivity. The GB plane within the straight segments is aligned normal to surface, as witnessed by the sharp appearing boundary projection using a top view TEM image (Fig. 7c). However, the GB plane within the curve exhibits several non-uniform inclinations which are evidenced in Fig. 7d,e by the Moiré pattern on both sides of the boundary, while the zone axes of both grains remains in the common <111> direction. The curvature spans over hundreds of nanometres and the projected width of the GB increases from 1–2 nm in the straight segment to 25 nm in the curved boundary (Fig. 7e). Therefore, a curved tilt GB plane is not aligned normal to surface and instead creates local spatial deviations as illustrated in Fig. 7f. The change in GB plane within a boundary curve yields less ordered atomic arrangements and even a possible loss of the tilt character of the GB. As a consequence, the resistivity increases by 80%. It must be noted that the increase in GB area leads to an underestimation of the GB resistivity due to the inverse relation between GB resistance and its area (equation 1), so the measured increase in GB resistivity due to curvature is a lower bound.

## Conclusions

In conclusion, this study provides an understanding of the relation between the GB structures and electrical resistivity. The research is based on direct and local resistivity measurements of a variety of [111] tilt sub-micron GB segments in Cu, accompanied with structural characterization by EBSD, TEM and MD simulations. GB resistivity is confirmed to depend on its CSL type, while it is not considerably affected by GB phases and defects within the same CSL type. The GB resistivity is correlated with the boundary's excess volume and excess energy in first approximation. The resistivities of LAGBs with misorientations of more than 10° are the highest among the tilt GBs, due to the high dislocation density and the resulting strain fields. GB curvature increases resistivity of the boundary due to deviation from tilt condition. Overall, this work provides a systematic experimental study on the impact of GB structure on its electrical properties.

## Methods:

### Thin film preparation and structural characterization

High purity Cu thin films (99.999% pure Cu) were deposited on (0001)-oriented α-$Al_2O_3$ substrates by magnetron sputtering at room temperature. The deposition was performed with a radio frequency (RF) power supply at 250W, 20 sccm Ar flow, and a background pressure of 0.66 Pa. Deposition time of 45 minutes yielded a nominally 600 nm thick film. Post-deposition thermal annealing was carried out at 400 °C for 2 hours within the sputtering chamber without breaking the vacuum. EBSD (EDAX detector in Zeiss Auriga SEM) analysis was employed to identify the crystallographic orientation of grains, as well as GB planes and type (OIM software). Subsequently, selected segments within individual GBs were selected based on the SEM-EBSD results and isolated for further investigation. The selected GB types are tilt GBs: Σ3, Σ7, Σ19b, Σ21a, Σ31a, Σ43b, and low angle GBs with $\theta = 7°$. The selected GB segments for each CSL type consist of different GB planes, deviations from ideal CSL angle and GBs with twist component. Table S1 shows the investigated GB segments. In addition, both macroscopically straight and curved segments were investigated.

The GB characteristics of each segment are evaluated by SEM-EBSD with an angular resolution of 0.5° and a step size of 30 nm. The reflections in the (111) pole figures from the adjacent grains yield their misorientation angle $\theta = (\theta_{CSL} + \delta\theta)$ and consequently determines the CSL type (see e.g. red and blue marks in Fig. 2d). The pole figures are also utilized for identifying the GB planes, using a suitable stereographic projection. The inverse pole figure plot readily reveals a possible out-of-plane misalignment of grains. GBs are confirmed to be aligned vertical to surface, with maximum inclination of 3°, as shown in Ref. [21].

**Grain boundaries segment fabrication and electrical measurements**

Electrical investigation of the chosen GB segments is detailed elsewhere. [21] Briefly, each GB segment was isolated from its surrounding film by milling trenches along the whole films' depth. The 30 μm long and ~ 0.5 μm wide trenches were created by focused ion beam (FIB-Zeiss Auriga) employing a beam current of 50 pA. *In-situ* SEM electrical characterization was conducted utilizing four probes provided by four needles having 50 nm tip radius, which are driven by four independent micromanipulators (Kleindiek – PS4). The resistivity measurements were done utilizing the direct current (dc) pulse method, with 5 mA pulse height and 10 ms pulse width.

**Molecular dynamics simulations**

MD simulations were performed with an embedded atom method (EAM) potential for copper [35] using LAMMPS. [47] First, grain boundaries were assembled from two appropriately-oriented crystallites with the ⟨111⟩ tilt axis in *z* direction and the GB normal in *y* direction. The bicrystals had a size of approximately $30 \times 20 \times 6$ nm³ (corresponding to around 300,000 to 400,000 atoms) with periodic boundaries in *z* direction and open boundaries otherwise. The open boundaries in contact with the GBs serve as reservoirs for interstitials and vacancies to allow diffusion-driven GB phase transformations. [24] These systems were annealed at 800 K for 4 ns and subsequently cooled to 300 K with a barostat at 0 Pa applied in *z* direction and an integration timestep of 2 fs. Unit cells of the GB phases were cut from these samples, made into cells with periodic boundaries in *x* and *z* direction, and scaled to fit the 0 K fcc lattice constant of the copper potential (3.615 Å). Atomic positions were then minimized with regard to the potential energy. The excess properties were calculated from these samples. [38] The structures were visualized with Ovito. [48]

**TEM**

Atomic structure of the GBs was analysed using aberration-corrected TEM and STEM, both operated at 300 kV (Titan Themis 60-300, Thermo Fisher Scientific). TEM samples were prepared by using site-specific plane-view lift-out method using a FIB-SEM dual beam workstation (Scios 2, Thermo Fisher Scientific). A 30 kV Ga ion beam was used for the cutting and rough milling and a 5 kV with 48 pA beam was used for the fine milling and cleaning. The probe current of 80 pA was used for high-resolution STEM HAADF imaging with a collection angle of 78-200 mrad and the convergence angle was 23.8 mrad. To minimize scan noise and specimen drift during the acquisition, rigid registration was applied averaging 10 frames recorded with dwell time of 1 µs. Strain distribution at the GB was calculated using GPA, which calculates relative changes in the lattice spacing based on FFT. The strain map in Fig. 6c is based on the reflections marked in the FFT in the inset

**Acknowledgement**

H.B, T.B and G.D. acknowledge the financial support by the ERC Advanced Grant GB CORRELATE (Grant Agreement 787446 GB-CORRELATE). S.L. acknowledge the financial support by Alexander von Humboldt Foundation.

# Supplemental information

**Table S1:**

List of inspected [111] tilt grain boundary (GB) segments for which GB resistivity was measured. In case of symmetric GBs only one of the GB normals is given since the GB normal of the other grain is identical.

| GB type | Misorientation (°) | GB plane normals | GB Resistivity ($10^{-12}$ $\Omega cm^2$) |
|---|---|---|---|
| Low angle GBs | 7 | [1 -4 3] ; [1 -8 7] | 0 ± 2 |
| | 13.8 | [1 -3 2] ; [1 -10 9] | 20.1 ± 1.5 |
| | 14 | [-10 -1 11] ; [-2 -1 3] | 19.7 ± 1.2 |
| | 16 | [1 -4 3] ; [ -2 1] | 22.9 ± 1.7 |
| | 16.6 | [3 -5 2] ; [7 -8 1] | 24.8 ± 2.8 |
| Σ 21a | 20 | [3 -1 -2] | 17.3 ± 3.5 |
| Σ 7 | 42.3 | [3 2 -5] ; [3 -1 -2] | 15 ± 1.7 |
| | 42.4 | [2 -5 3] ; [10 -9 -1] | 16.1 ± 2 |
| | 42.7 | [3 2 -5] ; [3 -1 -2] | 16 ± 2 |
| Σ 19b | 43.7 | [3 -1 -4] ; [2 -1 -1] | 10.4 ± 1.2 |
| | 44.4 | [1 0 -1] ; [4 -3 -1] | 10.2 ± 2.7 |
| | 45.5 | [2 -3 -1] ; [9 -5 -4] | 9.9 ± 2 |
| | 45.6 | [2 3 -5] | 12.8 ± 1.8 |
| | 47.5 | [1 -8 7] | 14.5 ± 4.5 |
| Σ 37c | 51.3 | [5 -1 4] ; [2 3 -5] | 7.3 ± 2 |
| | 52.1 | [3 1 -4] ; [5 -3 -2] | 6.7 ± 1.5 |
| | 53.3 | [1 3 -4] ; [-2 3 -1] | 5.4 ± 1.4 |
| Σ 3 | 58 | [1 -1 0] | 1.13 ± 0.26 |
| | 58.5 | [1 -1 0] | 1.25 ± 0.17 |
| | 59.5 | [1 -1 0] | 1.24 ± 0.1 |
| | 60 | [1 -1 0] | 1.18 ± 0.18 |

**Figure S1:**

Snapshots of the GB phases discovered using MD simulations viewed from the ⟨111⟩ direction. The two possible symmetric variants are shown in the left and right column, respectively. For Σ21a {123} and Σ19b {178}, two different phases can occur. Atoms inside the GB are highlighted in red to emphasize the structural motifs and guide the eye. They do not correspond to a structural analysis. The black bars indicate the GB unit cell. The scale bar applies to all images.

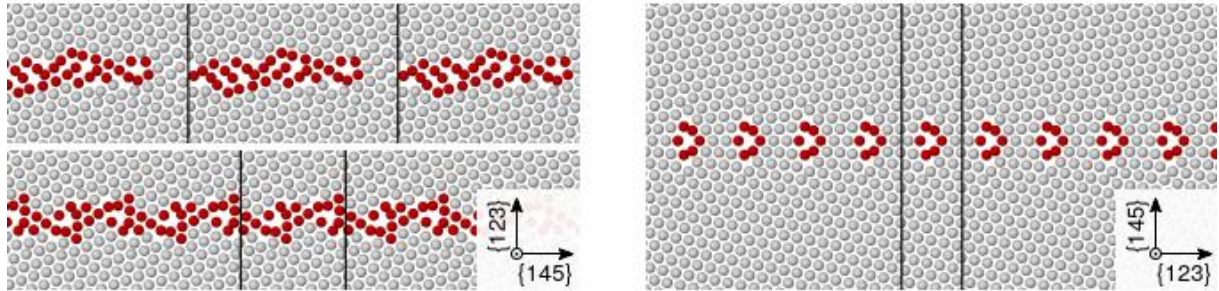

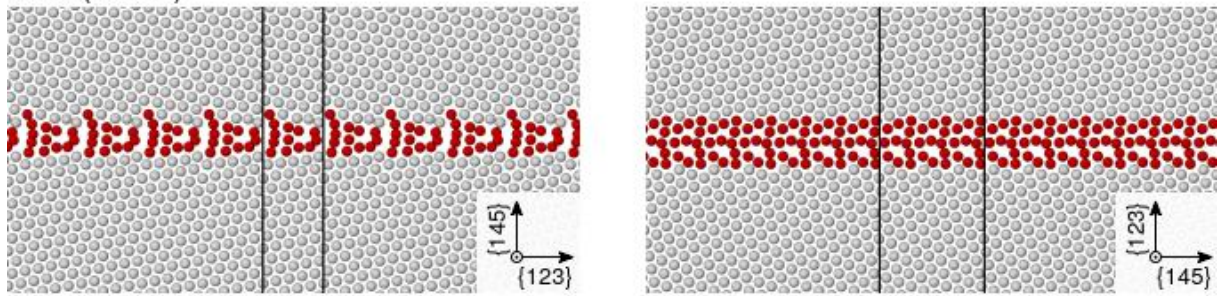

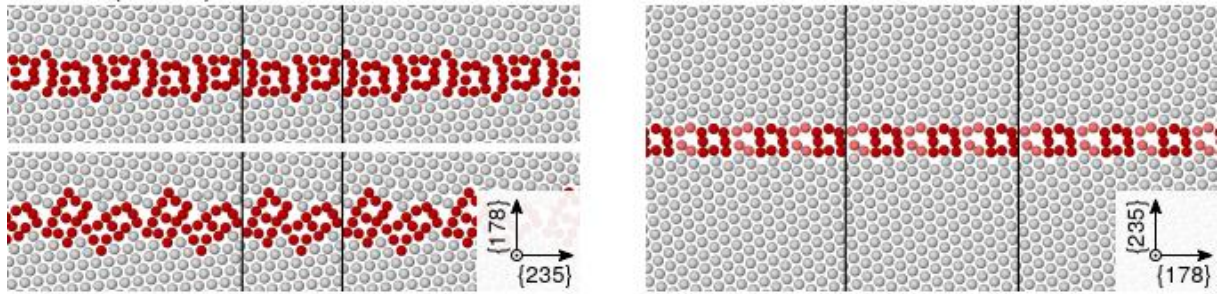

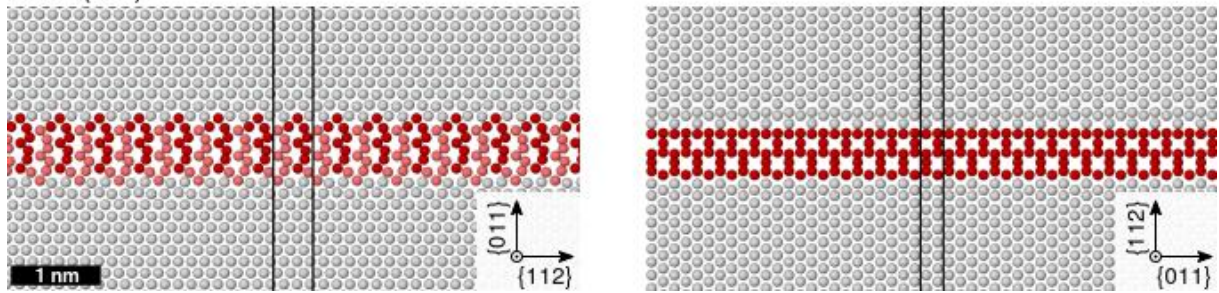